\documentclass[12pt,preprint]{aastex}

\def\gsim{\;\lower4pt\hbox{${\buildrel\displaystyle >\over\sim}$}\;}
\def\lsim{\;\lower4pt\hbox{${\buildrel\displaystyle <\over\sim}$}\;}
\def\grls{\;\lower4pt\hbox{${\buildrel\displaystyle >\over <}$}\;}

\shortauthors{WANG \& ZHANG} \shorttitle{Fast-CME ARs}

\begin{document}

\title{A Statistical Study on Solar Active Regions Producing Extremely Fast
Coronal Mass Ejections}

\author{Yuming Wang\altaffilmark{1, 2}, and Jie Zhang\altaffilmark{1}}
\altaffiltext{1}{Department of Computational and Data Sciences,
College of Science, George Mason University, 4400 University Dr.,
MSN 6A2, Fairfax, VA 22030, USA, Email: ywangf@gmu.edu,
jzhang7@gmu.edu}

\altaffiltext{2}{School of Earth \& Space Sci., Univ. of Sci. \& Tech. of
China, Hefei, Anhui 230026, China, Email: ymwang@ustc.edu.cn}

\begin{abstract}
We present a statistical result on the properties of solar source regions 
that have produced  57 fastest front-side coronal mass ejections (CMEs) 
(speed $\geq$1500 km s$^{-1}$) occurred from 1996 June to 2007 January. The 
properties of these fast-CME-producing regions, 35 in total,  are compared 
with those of all 1143 active regions (ARs) in the period studied. An 
automated method, based on SOHO/MDI magnetic synoptic charts, is used to 
select and characterize the ARs.  For each AR, a set of parameters 
are derived including the areas (positive, negative and total, denoted by 
$A_P$, $A_N$ and $A_T$ respectively), the magnetic fluxes (positive, 
negative and total, $F_P$, $F_N$ and $F_T$ respectively), the average magnetic 
field strength ($B_{avg}$), quasi-elongation ($e$) characterizing the overall 
shape of the AR, the number and length of polarity inversion lines (PILs, 
or neutral lines, $N_{PIL}$ and $L_{PIL}$ respectively), and the average and 
maximum of magnetic gradient on the PILs ($GOP_{avg}$ and $GOP_{max}$ 
respectively). Our statistical analysis shows a general trend between the 
scales of an AR and the likelihood of producing a fast CME, i.e., the larger 
the geometric size ($A_T$), the larger the magnetic flux ($F_T$), the stronger 
the magnetic field ($B_{avg}$), and/or the more complex the magnetic configuration
($N_{PIL}$ and $L_{PIL}$), then the higher the possibility of producing a fast 
CME. When all ARs are sorted into three equally-numbered groups with low, middle 
and high values of these parameters, we find that, for all these AR parameters, 
more than 60\% of extremely fast CMEs are from the high-value group. The two 
PIL parameters are the best indicators of producing fast CMEs, with more than 
80\% from the high value group.
\end{abstract}

\section{Introduction}

Coronal mass ejections (CMEs) are the most spectacular eruptive phenomenon in 
the solar corona,  which ejects a large amount of mass and magnetic flux into 
the interplanetary space and may cause adverse disturbances in the geo-space. 
From 1996 to 2007, more than 10 thousand CMEs have been observed by the LASCO 
(Large Angle and Spectrometric Coronagraph, \citet{Brueckner_etal_1995}) 
instrument on board SOHO (Solar and Heliospheric Observatory) spacecraft.  
Based on the CDAW CME catalog\footnote{http://cdaw.gsfc.nasa.gov/CME\_list/}, 
the CME speeds measured could be as low as $\sim21$ km s$^{-1}$ and as high as 
$\sim3387$ km s$^{-1}$, and the average speed is about 450 km s$^{-1}$. Among 
them, fast CMEs are of particular interest. Fast CMEs originate from solar 
active regions with a large amount of free magnetic energy that gives rise to 
the kinetic energy of a CME. Further, fast CMEs are usually much more 
geoeffective than slow CMEs, e.g, producing large solar energetic particle events 
and causing major geomagnetic storms.  In this paper, we focus on the extremely 
fast CMEs originating from the front-side solar disk, i.e., 
faster than 1500 km s$^{-1}$ or about 3 
times faster than  the average speed. These fast CMEs occupy a very small 
fraction ($\sim0.5\%$) of all CMEs.

The speed of a CME may be closely tied to the magnetic properties of its surface 
source region. While a CME could originate from a quiet Sun region with filament 
(or filament channel), it is believed that most CMEs, like most flares, originate 
from active regions (ARs), where strong magnetic field occupies a relatively 
large area. In a statistical study of 32 CME events, \citet{Subramanian_Dere_2001} 
found that 84\% CMEs were associated with ARs. \citet{Yashiro_etal_2005} suggested 
that almost all CMEs associated with sizable flares (flare larger than C3.0) 
occurred in ARs, and the percentage could reach up to 99\% (private communication). 
According to the flare catalogs from 2004 to 2007 during which the surface locations 
of almost all flares are listed (compiled by NOAA/SWPC, Space Weather Prediction 
Center), 79\% of all recorded flares originate from ARs, and the percentage rate is 
higher for larger flares. It is 89\% for C-class and above flares, and 98\% for M 
and X-class flares. Thus, the properties of ARs are the major concern in studying 
the production of CMEs and flares.

The relation  between flare productivity and AR properties has been studied 
extensively by many researchers \citep[e.g.,][]{Sammis_etal_2000, Leka_Barnes_2003b,
Leka_Barnes_2007, Maeshiro_etal_2005, Jing_etal_2006, Ternullo_etal_2006, 
Schrijver_2007, Georgoulis_Rust_2007}. In general, larger flares tend to occur in 
ARs with more complicated morphology, larger magnetic flux, larger magnetic energy, 
larger helicity, and stronger and longer neutral lines.

Nevertheless, CMEs are different phenomena from flares even though they are 
related in many aspects \citep[e.g.,][]{Harrison_1995, Harrison_2003,
Zhang_etal_2001a}.  The relations between CME productivity/properties and ARs 
properties have not been well pursued until recently. Examining 6 active regions 
with vector magnetogram observations, \citet{Falconer_etal_2006} showed a good 
correlation between AR magnetic properties (including the total magnetic flux, 
three other parameters related with main neutral lines, and two more parameters 
related with magnetic field twist) and CME productivity. \citet{Guo_etal_2007} 
investigated 55 flare-CME productive active regions with four magnetic parameters 
(tilt angle, total magnetic flux, length of main neutral lines and effective distance), 
and found that fast CMEs tend to occur in ARs with large magnetic flux and long 
length of main PILs.

In this paper, we  focus on the top 0.5\% fastest front-side CMEs and characterize their 
source ARs. Further, we use an automated method to select and characterize
all ARs observed by MDI from 1996 June to 2007 January, in order to create a 
matrix for comparison. The events and ARs studied cover almost the whole
solar cycle 23. The data gathering and processing methods are introduced
in the following section. The statistical results are presented in
section \ref{sec_results}. Section \ref{sec_summary} summarizes the work.

\section{Data and Methods}

\subsection{CME Selection and Source Region Identification}

\begin{figure}[tb]
  \centering
  \includegraphics[width=1.\hsize]{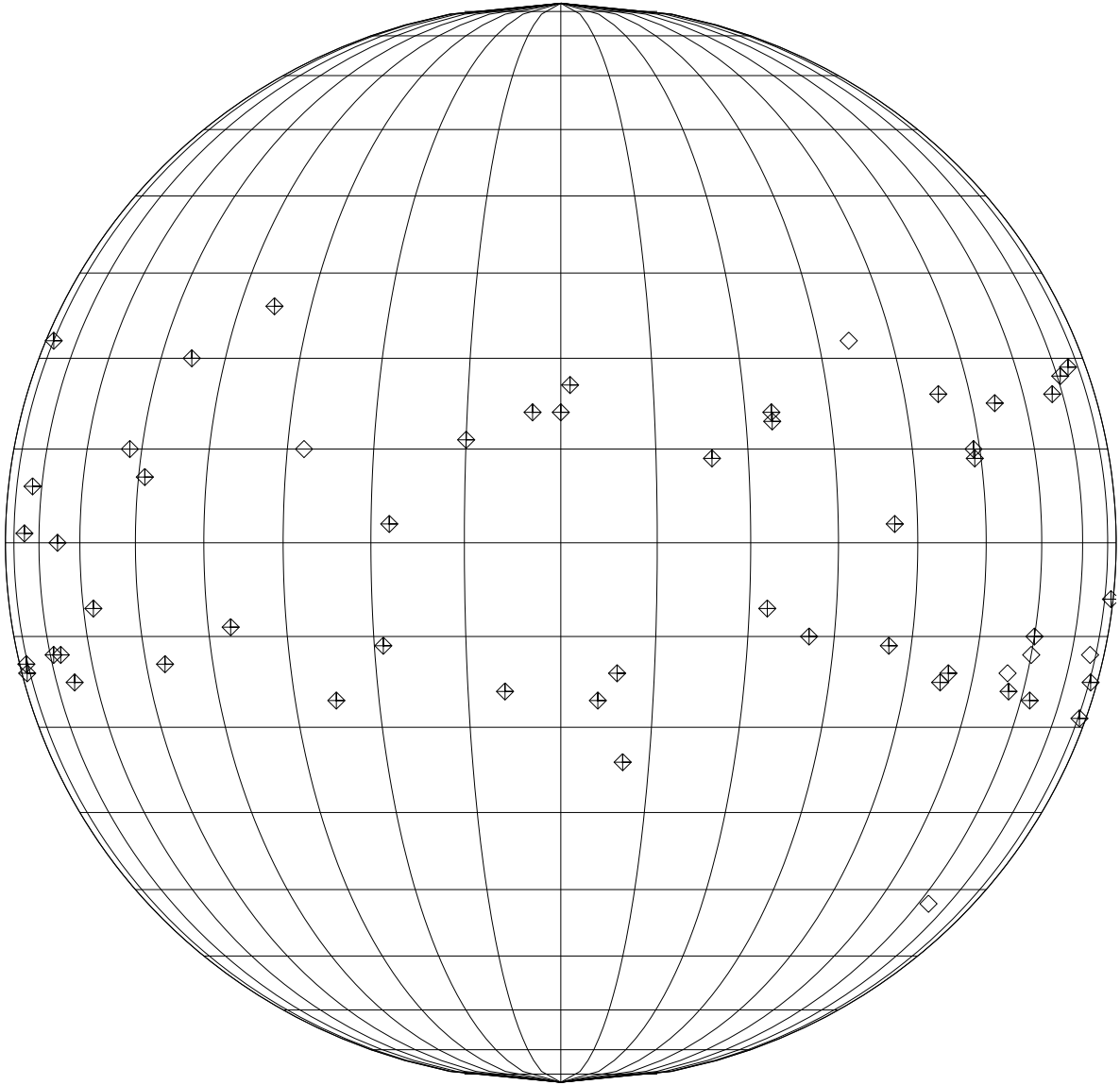}
  \caption{Distribution of heliographic coordinates of the surface source 
regions of the extremely fast CMEs. ARs are indicated in diamond-plus symbols 
and queit-Sun regions are indicated in diamond symbols.}\label{fg_cme_loc}
\end{figure}

Based on the CDAW CME catalog, the numbers of CMEs with speed over 
1000, 1500, 2000, and 2500 km s$^{-1}$ from 1996 June to 2007 January
are 496, 122, 37 and 9, respectively. For the sake of avoiding
the time-consuming manual examination of all fast CMEs, we choose the CME category of
speed $\geq1500$ km s$^{-1}$, which has a limited population of 122 events and
is more than 3 times faster than the CME average speed. Through visual 
examination of SOHO/EIT (EUV Imaging Telescope, \citet{Delaboudiniere_etal_1995}) 
and LASCO movies, 57 CMEs are found to originate from the front-side of solar 
disk that can be used to study the source regions. Table \ref{tb_cmes_ars} 
lists these front-side fast CMEs, including their speeds and source region 
locations. These CMEs (except the 55th and 56th CMEs which are determined 
according to the flare information in absence of EIT data) have definite 
eruptive signatures, including large scale dimming, compact brightening, 
and/or post-eruption loop arcades as seen in the SOHO/EIT corona images. 
Figure \ref{fg_cme_loc} shows the distribution of the heliographic coordinates 
of the surface source region of these CMEs. Apparently, the sample is uniform in
longitude, along which they are almost equally distributed; there is no bias 
toward the limbs, from which a CME may be faster due to smaller projection 
effect on speed measurement. 

\clearpage
\clearpage

\subsection{Determining the Properties of Active Regions}\label{sec_AR_method}

A catalog of solar ARs is compiled on a daily basis by NOAA/SWPC
\footnote{http://www.sec.noaa.gov/ftpmenu/forecasts/SRS.html}. It
is based on the active region report in near-real time from
approximately half a dozen ground-based observatories around the
globe, including observations of white light, $H_\alpha$ and
magnetogram. The compiled Solar Region Summary (SRS), for each
identified AR (or a sunspot group) on the front disk, lists the
assigned NOAA AR number, its heliographic coordinate (rotated to
24:00 UTC), Carrington longitude, total corrected area of the
sunspot group, Modified Zurich classification
\citep{McIntosh_1990}, longitudinal extent of the group in
heliographic degrees, total number of visible sunspots in the
group, and magnetic classification of the group. For the 11 year
period from 1996 to 2006 inclusive, there are in total 2999 active
regions listed by NOAA. Apparently, the NOAA AR catalog does not
provide the necessary quantitative information, such as magnetic
field strength and total magnetic flux, for the purpose of
studying the source regions of fast CMEs. In this paper, we have 
generated one version of our own AR catalog from the long term SOHO/MDI
(Michelson Doppler Imager) observations from 1996 to 2006.

To simplify the process of identifying and quantifying ARs, we
have used the Carrington rotation (CR) synoptic charts of MDI
magnetogram instead of individual magnetograms. When an AR
trans-crosses the front disk of the Sun in about 2 weeks, its
appearance may vary due to the intrinsic evolution and as well the
projection effect in the time-lapse magnetograms. On the other
hand, a synoptic chart is assembled from the stripe of data in the
central meridian from individual magnetograms observed over the
course of one Carrington rotation. By using synoptic charts the
projection effect is largely reduced. However, the possible
evolution of ARs is simply ignored. Since the launch of SOHO
spacecraft in Dec. 1995, high resolution synoptic charts of
photospheric magnetic fields from MDI are routinely made by the
MDI team at Stanford University
\footnote{http://soi.stanford.edu/magnetic/index6.html}. There are
143 synoptic charts from 1996 May to 2007 January, covering almost
the whole solar cycle 23.

We have developed an automated method to uniformly identify the ARs in MDI
synoptic charts. Attempts to detect ARs automatically have been made before,
for example, based on simple thresholdings \citep[e.g.,][]{Worden_etal_1998, 
Brandt_Steinegger_1998, Preminger_etal_2001}. \citet{Turmon_etal_2002} 
utilized pattern recognition techniques to distinguish different features on 
the solar surface. \citet{Benkhalil_etal_2006} used a more sophisticated 
approach by applying the region-growing segmentation technique in addition 
to thresholding. In this study, we use the region-growing method to automate 
the AR identification and characterization processes as described in the 
following.

(1) Find the kernel or seed pixels in the synoptic chart. We
choose the threshold of 15 times of the standard deviation,
$\sigma$, of the quite Sun region (typically $|B_{los}|\leq50$ G).
(2) Applying the morphological closing procedure. Isolated small
features, including anomaly noisy pixels, are effectively removed
by applying a morphology erosion operation followed by a
morphology dilation operation with a size of 15 pixels on the
$3600\times1080$ pixel map. (3) Applying the region-growing procedure.
Starting from the kernels defined above, all connected pixels with
$|B_{los}|\geq 5\sigma$ are included in the grown region. (4)
Applying the morphological opening procedure. Regions that are very close in space are
merged, by first applying the dilation operation followed by the
erosion operation with a size of 60 pixels. 

\clearpage
\begin{figure*}[tb]
  \centering
  \includegraphics[width=1.\hsize]{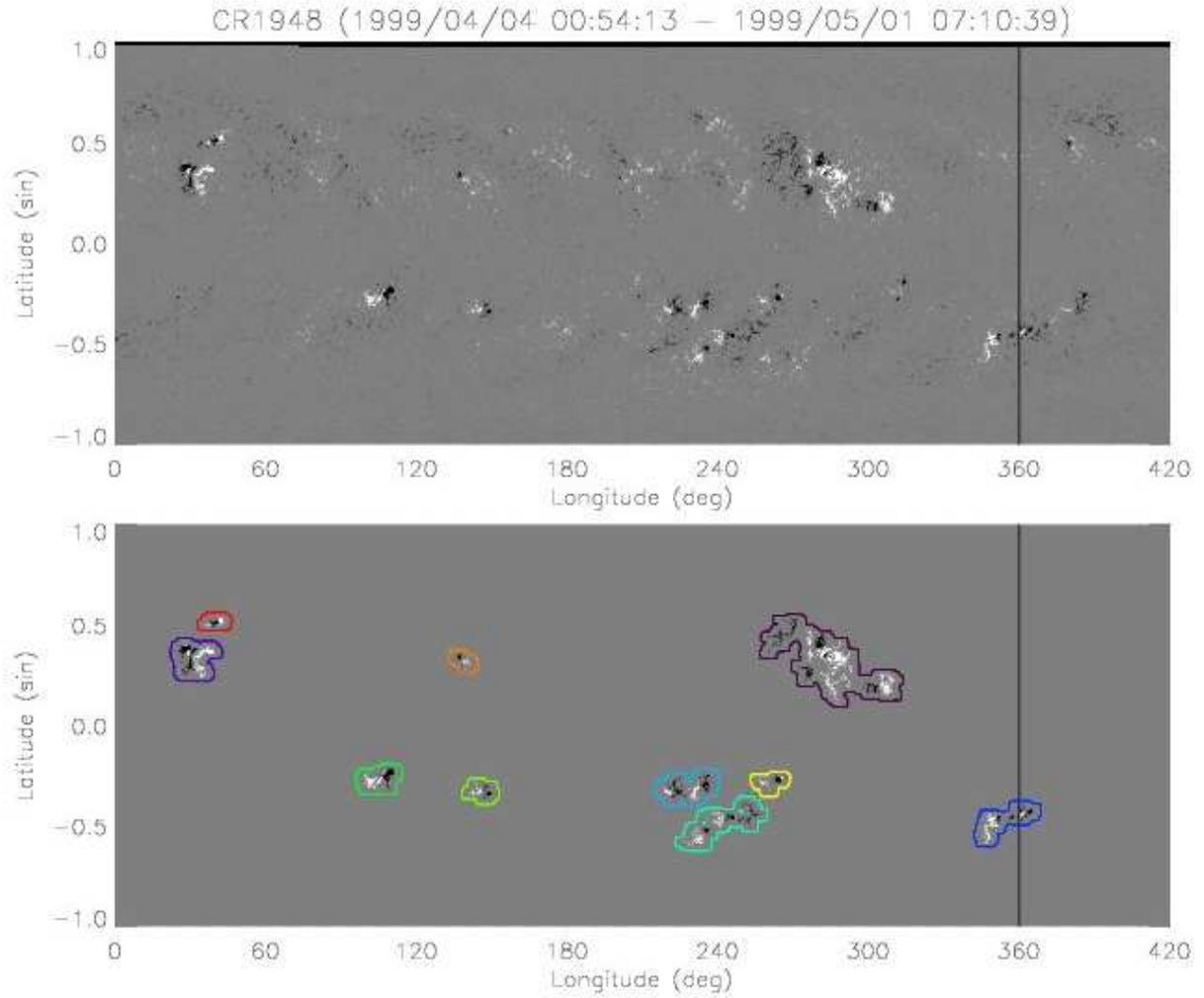}
  \caption{An example showing the original MDI synoptic chart (top) and 
the map showing only ARs extracted with our automated method (bottom).}
  \label{fg_ars_example}
\end{figure*}
\clearpage

Figure \ref{fg_ars_example} illustrates the result of the morphological analysis 
processes described above. The top panel shows the MDI synoptic chart of 
CR 1948 corresponding to the period from 00:54 UT on April 4, 1999 to 07:11 
UT May 1, 1999, and the bottom panel highlights all the extracted active
regions. Ten individual ARs are indicated by the lines with different colors.
The synoptic chart actually processed is an enlarged chart which is patched on 
the right-most with a left-most portion of the previous synoptic chart 
separated with the vertical black line; this treatment is to prevent those 
ARs crossing the Carrington rotation boundary from being cropped. The details 
of the imaging processing methods, including how sensitive the results depend 
on various intensity thresholds and area boxes chosen, will be pursued in another paper. 
Changing thresholds on AR selection will not have significant impact on the 
results of this paper, since this study mainly concerns the relative properties 
of ARs. 

\begin{figure}[tb]
  \centering
  \includegraphics[width=1.\hsize]{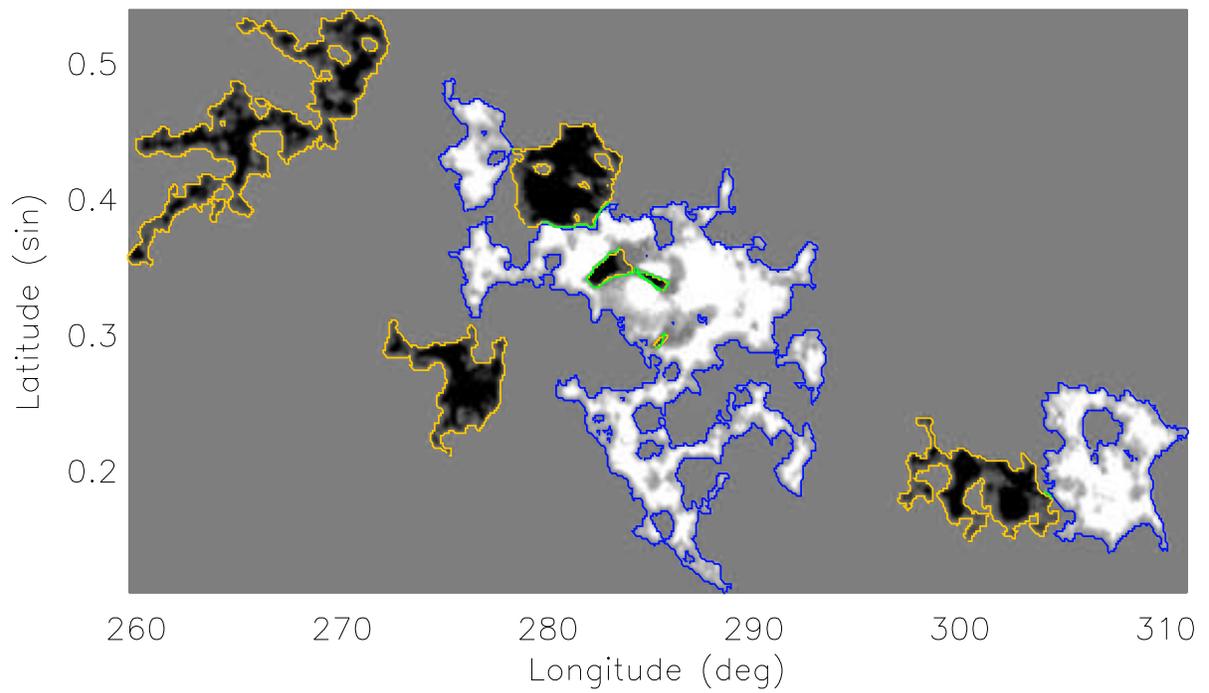}
  \caption{An AR of Carrington rotation of 1948 showing the positive (white
patches with blue boundaries), negative (black with yellow boundaries)
polarities and inversion lines (green curves) between them.}
  \label{fg_ar_example}
\end{figure}

Once an AR is identified and its boundary is determined, it is
straightforward to extract many useful parameters of the AR. We
use one AR in CR 1948 (Fig. \ref{fg_ar_example}) to illustrate the
characterization. The white and black patches, enclosed by the
blue and yellow lines, correspond to the regions of positive and
negative polarities in that active region. We calculated the area size
of the positive polarities ($A_P$), the size of the negative
polarities ($A_N$) and the total size ($A_T = A_P + A_N$). We also
calculated the total positive magnetic flux ($F_P$), total
negative magnetic flux ($F_N$) and the total unsigned flux
($F_T$). We derived the average magnetic field strength ($B_{avg}
= F_T/A_T$). We also defined a quasi-elongation parameter of ARs,
($e=1-\frac{D_s}{D_{max}}$, where $D_s$ is the radius of an assumed
circle of total size $A_T$ and $D_{max}$ is the largest distance
between two points in an AR). For the AR shown in Figure
\ref{fg_ar_example}, we found that the $A_P$, $A_N$, $A_T$, $F_P$,
$F_N$, $F_T$, $B_{ave}$ and $e$ are $1.9\times10^4$ Mm$^2$,
$1.2\times10^4$ Mm$^2$, $3.1\times10^4$ Mm$^2$,
$505.7\times10^{12}$ wb, $280.7\times10^{12}$ wb,
$786.4\times10^{12}$ wb, 72.8 G, and 0.32, respectively. These
parameters quantify the intensity and morphology of the AR, which
may be useful in relating ARs with eruptive phenomena.

\clearpage
\clearpage

Further, we extracted the polarity inversion lines (PIL, or
neutral lines) of the host AR. It is generally believed that the
properties of PILs are related with the stored free energy of the
AR studied. For each AR, we determined the number of PIL
($N_{PIL}$) and the  length of these PILs ($L_{PIL}$). The PILs of
AR are indicated by the green curves in Figure \ref{fg_ar_example}. 
 A PIL is a gathering of linearly linked pixels, which are  between pixels that have opposite
polarities. In this AR there are 6 separate PILs with a total
length ($L_{PIL}$) of 185.2 Mm. We also determined the average
magnetic gradient across these PILs ($GOP_{avg}$) and the maximum
gradient ($GOP_{max}$); the numbers are 152.7 and 496.4 G
Mm$^{-1}$ respectively.

In summary, a total of 12 magnetic parameters are derived. 
$A_T$ indicates the size of an AR; $F_T$ and $B_{avg}$ suggest 
how strong an AR is; $e$ indicate show much the overall shape of 
an AR deviates from a perfect circle; 
$N_{PIL}$ and $L_{PIL}$ indicate the morphological complexity of an AR; and 
$GOP_{avg}$ and $GOP_{max}$ is a measure of the free energy stored between 
two opposite polarities. These parameters are expected to have certain 
relations with the properties of eruptive events resulted. For instance, 
$F_T$ is ranked number one in the effectiveness of predicting flare occurrence 
from a comprehensive discriminant data analysis \citet{Leka_Barnes_2007}. A 
similar positive correlation between $F_T$ and large flare productivity was 
also obtained by \citet{Schrijver_2007}. Besides, ARs with larger
area and/or more complicated configuration tend to produce larger flares
\citep[e.g.,][]{Sammis_etal_2000, Ternullo_etal_2006}.

\subsection{Comparison with NOAA Active Regions}

\clearpage
\begin{figure}[tb]
  \centering
  \includegraphics[width=1.\hsize]{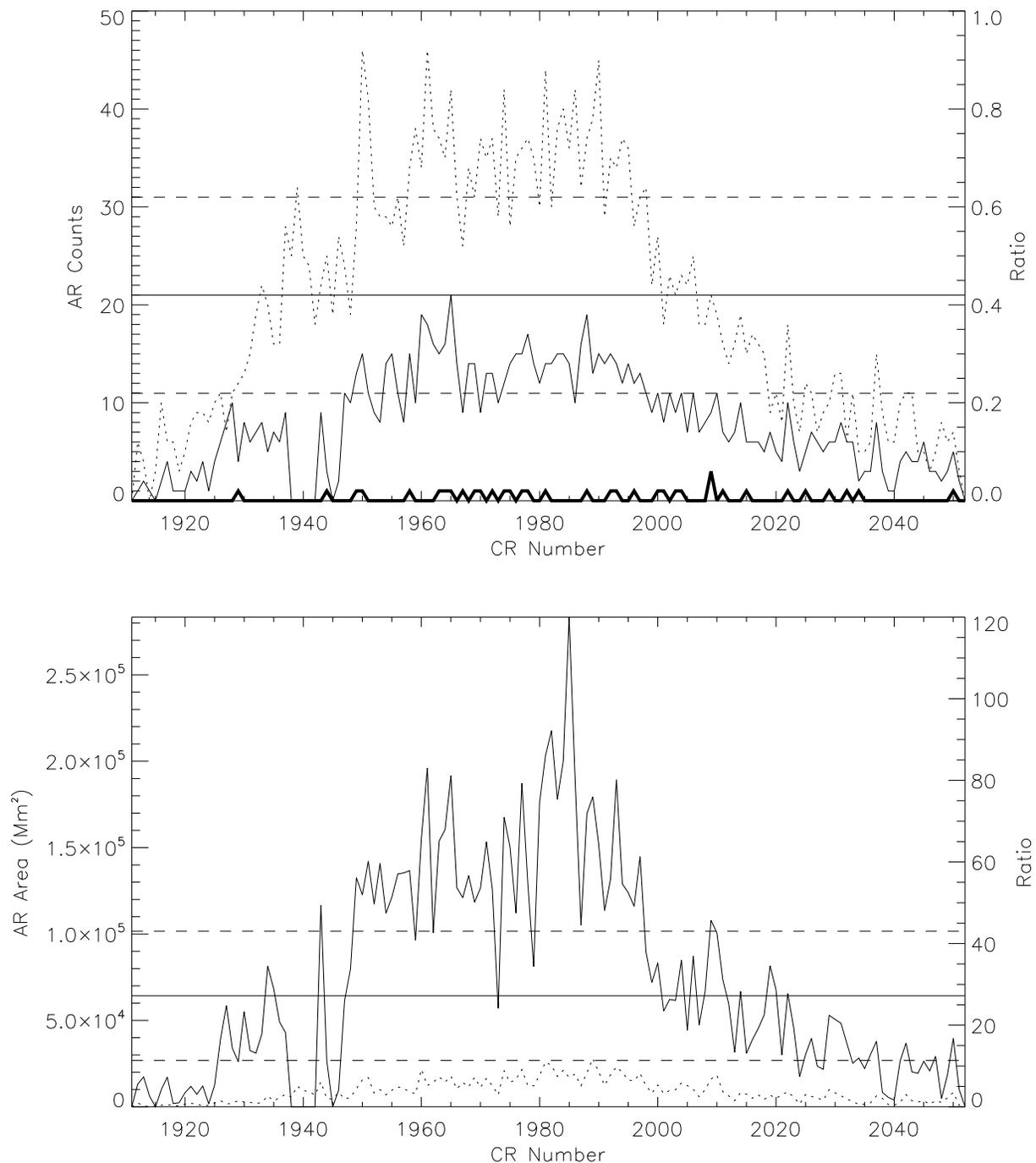}
  \caption{Comparison of our identified ARs with NOAA ARs from CR1911 to
CR2052 in terms of the count and area. Upper panel shows the AR counts and
the lower one shows the total AR area. The solid lines present our results
and the dotted lines NOAA results. The horizontal solid and dashed lines 
indicate the average ratio of our results to NOAA results and the standard 
deviations. The heavy solid line in the upper panel denotes the number of 
ARs hosting extremely fast CMEs.}
  \label{fg_comp_noaa}
\end{figure}
\clearpage

We have compared our MDI AR catalog with the NOAA AR catalog (Fig.
\ref{fg_comp_noaa}). The upper panel displays the AR counts of
each CR from our catalog (solid line) and NOAA catalog(dotted line). The
variation of the counts  along the solar cycle seems the same for the two
lines: small near the solar minimum and reaching the peak near the
solar maximum. However, the numbers of our ARs are smaller than that of
NOAA ARs. The ratio of the two counts is 0.42 averaged over the entire 
period studied with a standard deviation of 0.20. The smaller
number of ARs identified by our method is attributed
to several selection effects.(1) The usage of magnetogram data. 
Regions that are sufficiently close to each 
other in the distribution of magnetic field may be grouped into a 
single large AR. The selection of NOAA ARs is largely based on the visual appearance of 
sunspots, which are usually compact in size and discrete in distribution. On 
the other hand, the surface magnetic field, which we use to define the 
size, may extend to a much larger size than that of the sunspots. (2) 
ARs with weak magnetic fields are not selected because of the adoption of a high kernel 
threshold. (3) Some short lived ARs (emerged from the western 
hemisphere and lasting less than one Carrington rotation) are not 
caught in synoptic charts. A full comparison study between our catalog and 
the NOAA catalog will be made in a separate paper.

The lower panel of Figure \ref{fg_comp_noaa} presents the total AR
areas during each CR for both MDI ARs we identified and the NOAA
ARs. In contrast to the AR counts, the areas from our method
are much larger than those from NOAA. The ratio suggests that the
total AR area in our method is on average 27.2 times that by NOAA ARs.
The AR area estimated by NOAA concerns only sunspots, which
usually correspond to the regions with extremely strong magnetic fields. On
the other hand, we include all regions with $|B|\geq5\sigma$
connecting to the AR kernels.

\section{Results}\label{sec_results}

\begin{figure*}[tb]
  \centering
  \includegraphics[width=1.\hsize]{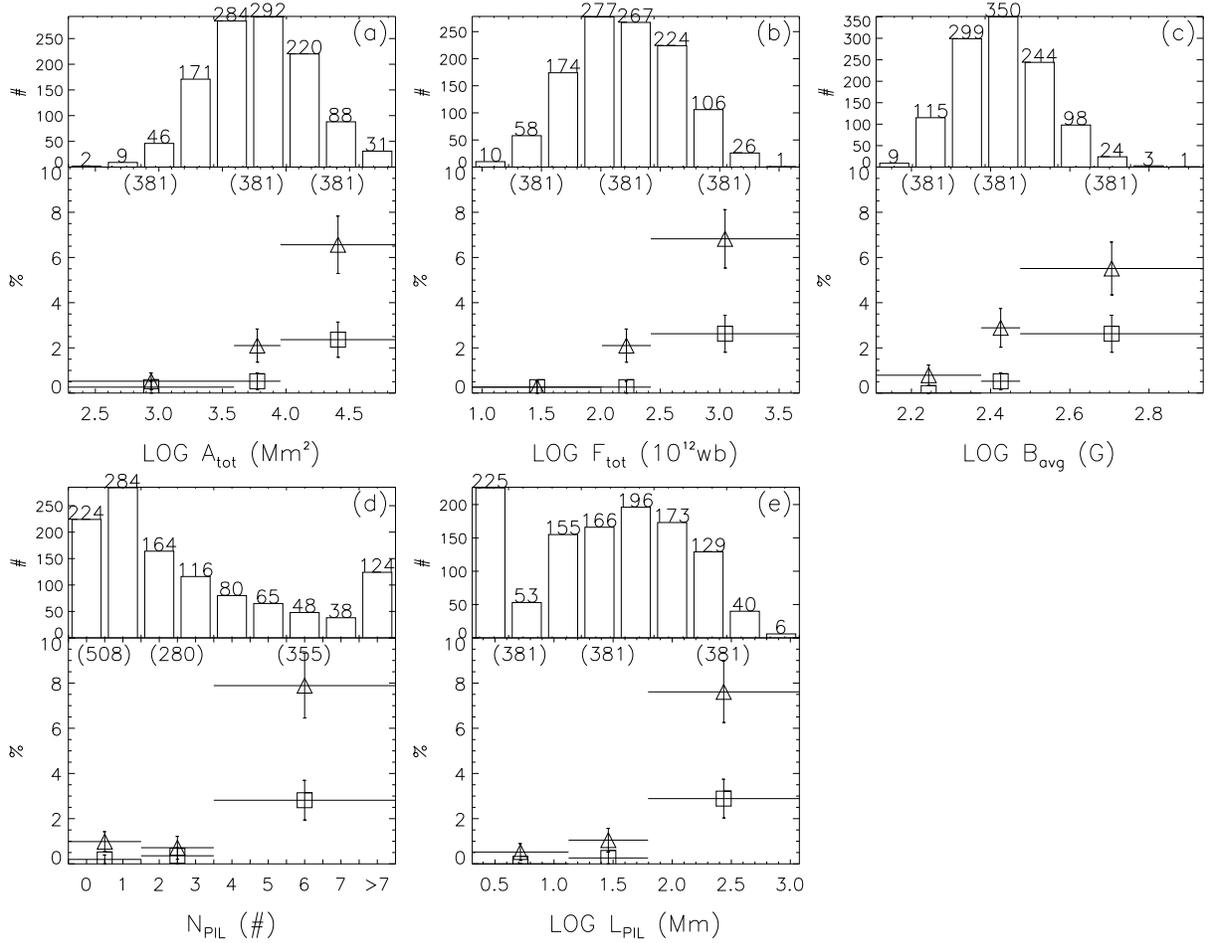}
  \caption{Histogram distributions of AR numbers on various AR parameters: 
(a) total area, (b) total magnetic flux, (c) average magnetic field, (d) 
number of PILs, and (e) total length of PILs. The first three and the last
parameters are plotted in log-scale along the x-axis. In each panel, the 
upper sub-panel shows the histogram for all ARs in nine bins, while the lower 
sub-panel shows the ratios of the ARs hosting extremely fast CMEs to all ARs 
in three equally-numbered general AR groups (see text for details). The 
numbers in parentheses give the number of ARs in the three groups. The 
symbols of triangles are for the extremely fast CMEs and 
the squares are for the fast CMEs originated between longitude of $\pm30^\circ$.} 
  \label{fg_ars_histo}
\end{figure*}

Using our method, we have identified and characterized 1143 MDI ARs 
for Carrington rotations 1911 to 2051 covering the period from 1996
June 28 to 2007 January 8. It seems that the 5 parameters that
characterize the overall scales of an AR ($A_T$, $F_T$, $B_{avg}$, 
$N_{PIL}$ and $L_{PIL}$) are all fairly related with the production of  fast CMEs. 
The histogram distributions of AR numbers of these five parameters are
shown in the five panels in Figure \ref{fg_ars_histo} (upper panel),
respectively. All parameters except $N_{PIL}$ are plotted in
log-scale and divided into nine bins. One can find that all the
distributions, except for $L_{PIL}$, are unimodal with a peak appearing 
at a certain value in between. The 
histogram (a) and (b), which are for $A_T$ and $F_T$, indicate the 
distributions of the geometric size and magnetic flux sizes of ARs, 
respectively. Most ARs have a size of $\sim10^{3.8}$ Mm$^2$ with the 
total magnetic flux peaked at $\sim10^{14}$ wb. This kind of 
distribution with a central peak is much different from that of the 
distribution of the size of sunspot groups studied before, which was
exponential \citep[e.g.,][]{Schrijver_1988, Howard_1996} or following
a power law \citep[e.g.,][]{Harvey_Zwaan_1993}. A detailed analysis on 
the causes of these differences is beyond the scope of this paper and 
will be pursued separately. In this paper, we focus on the ARs hosting 
extremely fast CMEs.

Table \ref{tb_cmes_ars} lists all the fast CMEs studied, the heliographic 
coordinates of the surface source regions and the NOAA AR numbers of the 
host ARs. 55 of the 57 events studied were originated from NOAA ARs. The 
rest two CMEs, numbered 2 and 20 in the table, were not originated in any 
NOAA ARs, nor in our MDI ARs. Careful inspection showed that these two 
fast CMEs were associated with the eruption of giant filaments. Four more 
CMEs (numbered 6, 23, 51 and 52) were not originating from our MDI ARs. 
The surface source regions of these CMEs are too small and/or weak in the 
MDI synoptic charts that they do not qualify to be an MDI AR using our 
method. A summary is given in Table \ref{tb_ar_cme_number}. In total, 	
89.5\% (51 out of 57) fast CMEs 
were originated from our MDI ARs (the locations of these CMEs have been 
indicated in diamond symbol with plus sign in Figure \ref{fg_cme_loc}). The
percentage of AR origin is as high as 96\% when NOAA ARs are used. 

These fast CMEs were originated from 35 individual MDI ARs. Compared 
with a total number of 1143 MDI ARs during the period studied, the 
percentage of ARs hosting these extremely fast CMEs is considerably small. 
It is only about 3.1\%. The
distribution with solar cycle of these fast-CME-hosting ARs has been 
plotted in the thick solid line in Figure \ref{fg_comp_noaa}. Apparently, 
the majority of these ARs occurred during the solar maximum.

\clearpage

\clearpage
\thispagestyle{empty}
\setlength{\voffset}{-15mm}
\begin{table*}[t]
\linespread{1.5} \caption{Front-side fast CMEs\tablenotemark{*} in solar cyle 23 and their surface source regions}
\label{tb_cmes_ars}
\tabcolsep 2pt
\footnotesize 
\begin{tabular}{r|ccccc||r|ccccc}
\hline
No.& Date Time\tablenotemark{a}& $V_{CME}$\tablenotemark{b}& Location& AR$_{NOAA}$ &AR\tablenotemark{c} & No.& Date Time\tablenotemark{a}& $V_{CME}$\tablenotemark{b}& Location& AR$_{NOAA}$ & AR\tablenotemark{c}\\
\hline
 1& 1997/11/06 12:10:41& 1556& S17W62&  8100& 1     &    30& 2003/06/17 23:18:14& 1813& S07E58& 10386& 24      \\
 2& 1998/04/20 10:07:11& 1863& S42W63&   -  & N     &    31& 2003/10/26 17:54:05& 1537& N02W37& 10484& 25      \\
 3& 1998/12/18 18:09:47& 1749& N26E35&  8415& 2     &    32& 2003/10/28 11:30:05& 2459& S16E06& 10486& 26      \\
 4& 1999/05/03 06:06:05& 1584& N20E45&  8525& 3     &    33& 2003/10/29 20:54:05& 2029& S17W04& 10486& 26      \\
 5& 1999/06/04 07:26:54& 2230& N19W75&  8552& 4     &    34& 2003/11/02 17:30:05& 2598& S16W57& 10486& 26      \\
 6& 2000/01/06 07:31:24& 1813& N22W34&  8816& N     &    35& 2003/11/04 19:54:05& 2657& S19W81& 10486& 26      \\
 7& 2000/05/15 08:50:05& 1549& N22E85&  9002& 5     &    36& 2003/11/18 08:50:05& 1660& N02E18& 10501& 27      \\
 8& 2000/06/25 07:54:13& 1617& N15W54&  9046& 6     &    37& 2004/01/07 04:06:07& 1581& N01E75& 10537& 28      \\
 9& 2000/07/14 10:54:07& 1674& N17W01&  9077& 7     &    38& 2004/01/07 10:30:29& 1822& N06E73& 10537& 28      \\
10& 2000/09/12 11:54:05& 1550& S19W06&  9163& 8     &    39& 2004/01/08 05:06:05& 1713& N00E65& 10537& 28      \\
11& 2000/11/08 23:06:05& 1738& N18W71&  9213& 9     &    40& 2004/04/11 04:30:06& 1645& S15W45& 10588& 29      \\
12& 2000/11/25 01:31:58& 2519& N07E49&  9240& 10    &    41& 2004/11/07 16:54:05& 1759& N09W16& 10696& 30      \\
13& 2001/01/20 21:30:08& 1507& S09E37&  9313& 11    &    42& 2004/11/09 17:26:06& 2000& N09W49& 10696& 30      \\
14& 2001/04/02 22:06:07& 2505& N16W67&  9393& 12    &    43& 2004/11/10 02:26:05& 3387& N10W49& 10696& 30      \\
15& 2001/04/10 05:30:00& 2411& S24W07&  9415& 13    &    44& 2005/01/15 06:30:05& 2049& N14E03& 10720& 31      \\
16& 2001/06/11 04:54:05& 1647& S13E81&  9501& 14    &    45& 2005/01/15 23:06:50& 2861& N14W00& 10720& 31      \\
17& 2001/07/19 10:30:22& 1668& S05W60&  9537& 15    &    46& 2005/01/17 09:30:05& 2094& N14W23& 10720& 31      \\
18& 2001/09/24 10:30:59& 2402& S17E25&  9632& 16    &    47& 2005/01/17 09:54:05& 2547& N13W23& 10720& 31      \\
19& 2002/04/21 01:27:20& 2393& S15W81&  9906& 17    &    48& 2005/01/19 08:29:39& 2020& N16W45& 10720& 31      \\
20& 2002/05/22 03:50:05& 1557& S14W56&   -  & N     &    49& 2005/05/13 17:12:05& 1689& N11E10& 10759& 32      \\
21& 2002/07/23 00:42:05& 2285& S12E69& 10039& 18    &    50& 2005/07/30 06:50:28& 1968& N10E52& 10792& 33      \\
22& 2002/08/16 12:30:05& 1585& S11E19& 10069& 19    &    51& 2005/08/22 17:30:05& 2378& S12W60& 10798& N       \\
23& 2002/09/05 16:54:06& 1748& N10E28& 10102& N     &    52& 2005/08/23 14:54:05& 1929& S12W77& 10798& N       \\
24& 2002/11/09 13:31:45& 1838& S10W27& 10180& 20    &    53& 2005/09/09 19:48:05& 2257& S12E67\tablenotemark{d}& 10808\tablenotemark{d}& 34\\
25& 2002/11/10 03:30:11& 1670& S11W37& 10180& 20    &    54& 2005/09/10 21:52:07& 1893& S13E47\tablenotemark{d}& 10808\tablenotemark{d}& 34\\
26& 2003/03/18 12:30:05& 1601& S14W46& 10314& 21    &    55& 2005/09/11 13:00:53& 1922&       & 10808\tablenotemark{e}& 34    \\
27& 2003/03/23 12:06:05& 1505& S13E71& 10318& 22    &    56& 2005/09/13 20:00:05& 1866&       & 10808\tablenotemark{e}& 34    \\
28& 2003/06/02 00:30:07& 1656& S06W85& 10365& 23    &    57& 2006/12/13 02:54:04& 1774& S07W22& 10930& 35        \\
29& 2003/06/15 23:54:05& 2053& S14E82& 10386& 24    &                                                          \\
\hline
\end{tabular}
\tablenotetext{a}{First appearance of CMEs in the field of view of LASCO/C2.}
\tablenotetext{b}{Linear speed of CMEs in the combined fields of view of LASCO/C2 and C3, in units of km s$^{-1}$.}
\tablenotetext{c}{Whether or not a CME originated from a MDI AR we identified. The sequenced numbers indicate a  AR, and `N¡¯ indicates negative association with AR}
\tablenotetext{d}{Information is based on the NOAA/SEC flare list.}
\tablenotetext{*}{Date, time and $V_{CME}$ are adopted from the CDAW LASCO CME Catalog.}
\end{table*}

\clearpage
\setlength{\voffset}{0mm}

\begin{table}[t]
\linespread{1.5} \caption{Statistical Numbers of extremely fast CMEs and the host active regions}
\label{tb_ar_cme_number}
\footnotesize 
\begin{tabular}{ll}
\hline
Number of all ARs&         1143\\
\multicolumn{2}{c}{For all the front-side fast CMEs}\\
Number of front-side CMEs&           57 \\
Number of front-side CMEs originating from ARs&           51 (89.5\%)\\
Number of ARs hosting these CMEs&          35 ($3.1\%$)\\
\multicolumn{2}{c}{For the fast CMEs originating from E30$^\circ$ to W30$^\circ$}\\
Number of front-side CMEs&          17\\
Number of front-side CMEs originating from ARs&           16\\
Number of ARs hosting these CMEs &           12\\
\hline
\end{tabular}
\end{table}
\clearpage

The overall distributions of these fast-CME-hosting ARs are shown along with 
that of all ARs in Figure \ref{fg_ars_histo}. While the distributions of all 
ARs (upper panel) are shown in 9 bins thanks to the large number, the 
distributions of fast-CME-hosting ARs are displayed in only three groups with 
low, middle and high values of the derived parameters, because of the limited 
number of ARs. The minimum-to-maximum values of the parameters in each group 
are such chosen that there are equal number of general ARs in each of three 
groups. The values used for such equal divisions are listed in the 2nd and 3rd 
columns in Table \ref{tb_AR_percentage}. Note that for $N_{PIL}$ the numbers can not be 
exactly equal, but are most close to each other, because $N_{PIL}$ is highly discrete.
Then we determine, in each of the three groups, the number of ARs producing 
fast CMEs, which indicate the effectiveness of the parameters used in discriminating 
fast-CME-hosting ARs from general population. The triangles shows the ratios of ARs 
hosting fast CMEs to all ARs, i.e., the probability of an AR producing fast CMEs. 
The horizontal bars indicate the value range of the groups, and the vertical 
bars indicate the uncertainty of the calculated percentage probability, defined 
as $\sqrt{\frac{p(1-p)}{N}}$, where $p$ is the probability and $N$ is the total 
sample number; this uncertainty is based on the assumption that the occurrence 
of fast CMEs obeys a binomial distribution.

It is obvious that the probability of producing fast CMEs increases as the 
value of the AR parameters increases. This trend is true for all the 
five parameters plotted in the Figure. The differences between low/middle and
high value groups all exceed the estimated uncertainty. In particular, 
$N_{PIL}$ and $L_{PIL}$ manifest the largest increase from low/middle to high
value groups; the probability increases by a factor as large as 8. This 
kind of trend indicates that the larger the AR parameters, i.e., the larger 
the geometric size, the larger the magnetic flux, the stronger the magnetic field, 
and the more complex the magnetic configuration, the higher the likelihood of it 
producing fast CMEs.

\clearpage
\begin{table*}[t]
\linespread{1.5} \caption{Percentages of fast-CME-hosting ARs in three different value groups. }
\label{tb_AR_percentage}
\tabcolsep 5pt
\footnotesize 
\begin{tabular}{c|cc|ccc|ccc}
\hline
Quantity& \multicolumn{2}{c|}{Separations}& $n^{low}$& $n^{mid}$& $n^{hig}$& $n^{low}_{cen}$& $n^{mid}_{cen}$& $n^{hig}_{cen}$\\
\hline
   $A_{tot}$&  3.90\tablenotemark{a}& 9.05\tablenotemark{a}&  2 ( 5.7\%)&  8 (22.9\%)& 25 (71.4\%)&  1 ( 8.3\%)&  2 (16.7\%)&  9 (75.0\%)  \\
   $F_{tot}$& 102.1\tablenotemark{b}&263.1\tablenotemark{b}&  1 ( 2.9\%)&  8 (22.9\%)& 26 (74.3\%)&  1 ( 8.3\%)&  1 ( 8.3\%)& 10 (83.3\%)  \\
   $B_{avg}$&   238\tablenotemark{c}&  298\tablenotemark{c}&  3 ( 8.6\%)& 11 (31.4\%)& 21 (60.0\%)&  0 ( 0.0\%)&  2 (16.7\%)& 10 (83.3\%)  \\
   $N_{PIL}$&     1&        3&      5 (14.3\%)&  2 ( 5.7\%)& 28 (80.0\%)&  1 ( 8.3\%)&  1 ( 8.3\%)& 10 (83.3\%)  \\
   $L_{PIL}$&  13.3\tablenotemark{d}& 62.4\tablenotemark{d}&  2 ( 5.7\%)&  4 (11.4\%)& 29 (82.9\%)&  0 ( 0.0\%)&  1 ( 8.3\%)& 11 (91.7\%)  \\
\hline
\end{tabular}
\tablenotetext{a}{in units of $10^3$ Mm$^2$.}
\tablenotetext{b}{in units of $10^12$ wb.}
\tablenotetext{c}{in units of Gauss.}
\tablenotetext{d}{in units of Mm.}
\tablenotetext{e}{in units of G Mm$^{-1}$.}
\end{table*}
\clearpage

Table \ref{tb_AR_percentage} lists the occurrence numbers and 
percentages of fast-CME-hosting ARs in the three groups discussed 
earlier (the fourth through sixth column). For all the parameters studied, more than 60\% 
fast-CME-hosting ARs lie in the high value group. The length of PIL 
($L_{PIL}$) has the largest percentages ($82.9\%$). These results indicate 
that the PIL-related parameters are the best in determining whether an AR 
would produce extremely fast CMEs.

As mentioned earlier, the evolution of ARs has not been taken into account when 
the AR parameters are derived, since the calculations were based on synoptic 
charts in order to minimize the projection effect. An AR may have evolved to 
some extent between the time of the CME occurrence and the time of 
central-meridian crossing; the time difference may be up to 7 days for those 
CMEs originating from the limbs. To find out how significant the temporal 
evolution affects our statistical results, we made a similar statistical study 
but considering only those CMEs that are close to the central meridian. The 
squares in Figure \ref{fg_ars_histo} and the last three columns, named 
$n_{cen}^{low}$, $n_{cen}^{mid}$ and $n_{cen}^{hig}$, in Table \ref{tb_AR_percentage} 
show the results for these near-central meridian CMEs (longitude 
between $\pm30^\circ$). It is found that there is no 
significant difference from those of all front-side CMEs. 

Some studies showed that flux variations (e.g., emergence or cancellation) 
are related with the CME initiation~\citep[e.g.,][]{Feynman_Martin_1995, 
Lara_etal_2000, Green_etal_2003, Sterling_etal_2007}. We check the Hale 
classes of all the corresponding NOAA ARs producing those fast CMEs studied. 
During the interval between the CME launch time and the central-meridian 
crossing of the corresponding source AR, it is found that 15 out of 35 ARs 
changed from one Hale class to another. However, for these 15 ARs with 
changes, nine are located very close to the limb when CMEs erupted, 
indicating a maximal projection effect which may affect the AR classification. 
Therefore, most of the ARs interested probably did not changed their Halo 
classes. While it may be related with CME initiation, the small scale 
flux variation may not alter the overall structure (and the stored energy) 
of the coronal magnetic field where CMEs originate. Considering that this 
study is a rather coarse quantitative statistical work, we believe that 
the AR evolution effect, when fully addressed, would not change our results.

\section{Summary}\label{sec_summary}

By applying the region-growing segmentation method, a total of 1143 ARs are 
extracted from 141 MDI synoptic charts covering the period from 1996 Jun 28 
to 2007 January 8. Twelve quantities ($A_P$, $A_N$, $A_T$, $F_P$, $F_N$, $F_T$, 
$B_{avg}$, $e$, $N_{PIL}$, $L_{PIL}$, $GOP_{avg}$ and $GOP_{max}$) are derived 
from the photospheric magnetic field distribution, concerning the size, strength, morphology,
complexity and free energy of ARs. These AR regions produced more than 10000 CMEs,
among which 122 CMEs were faster than or equal to 1500 km s$^{-1}$. We 
studied 57 of these extremely fast CMEs that originated from front-side ARs. 
Through a comparison of these fast-CME-hosting ARs with all other ARs, the 
following conclusions are reached.

(1) 89.5\%(51 out of 57) of fast CMEs are originated from our MDI ARs.
The number of ARs hosting these fast CMEs is 35, occupying 3.1\% of 
all ARs. A majority of these ARs occurred during the solar maximum.

(2) The distributions of the parameters, except $L_{PIL}$, are unimodal with a peak 
at a certain middle value. A typical MDI AR has a geometric size of $\sim10^{3.8}$ 
Mm$^2$ and a  magnetic flux of $\sim10^{14}$ wb.

(3) The derived AR parameters, $A_T$, $F_T$, $B_{avg}$, $N_{PIL}$ and $L_{PIL}$, 
all showed a positive correlation with the probability of an AR producing a
fast CME. When all  ARs are equally divided into three groups with low, middle and high 
value of the parameters studied, we find that the majority of fast-CME-hosting ARs 
resides in the high-value groups, e.g., the PIL-related parameters ($N_{PIL}$ and 
$L_{PIL}$) are particularly effective in indicating the occurrence of fast CMEs; 
the percentage can be as high as 82.9\%. Compared with the low-to-middle-value AR 
groups, the probability of producing a fast CME from a high-value AR can increase 
by a factor as large as 8.  These results suggest 	
that the larger, the stronger, and/or the more complex an AR, the more likely it 
produce an extremely fast CME.

The ARs hostingfast CMEs occupy a small fraction of all ARs 
($\sim3.1\%$). The fraction is about 8\% even for only the high-value AR groups.  
It indicates that there are many ARs having large-value parameters, but never 
producing an extremely fast CME. The current method is 
based on photospheric magnetic field observations only. However, we believe that
coronal magnetic fields play a direct role in determining the dynamic or 
kinematic properties of CMEs. In a statistical study of 99 halo 
CMEs, \citet{Liu_2007} suggested that CMEs under unidirectional open field 
structures are significantly faster than those under the heliospheric current 
sheet. To develop a better capacity of predicting the occurrence of fast CMEs, one 
need to study the AR parameters from both the photosphere and the corona as well.

\acknowledgments
We acknowledge the use of the solar data from the
MDI, LASCO, and EIT instruments on board SOHO spacecraft. The
SOHO/LASCO data used here are produced by a consortium of the
Naval Research Laboratory (USA), Max-Planck-Institut fuer
Aeronomie (Germany), Laboratoire d'Astronomie (France), and the
University of Birmingham (UK). SOHO is a project of international
cooperation between ESA and NASA. We also acknowledge the use of
CME catalog generated and maintained at the CDAW Data Center by
NASA and The Catholic University of America in cooperation with
the Naval Research Laboratory, and the flare list compiled by the
Space Environment Center of NOAA. This work is supported by
NSF SHINE grant ATM-0454612 and NASA grant NNG05GG19G. Y. Wang is
also supported by the grants from NSF of China (40525014) and
MSTC (2006CB806304), and J. Zhang is also supported by NASA
grants NNG04GN36G.

\bibliographystyle{kluwer}
\bibliography{../../ahareference}

\end{document}